# Evaluating the Utility of Research Articles for Teaching Information Security Management

**Harry Zurita**

Department of Computing and Information Systems
The University of Melbourne
Parkville, Victoria, Australia
Email: harry.zurita@gmail.com

**Sean B. Maynard**

Department of Computing and Information Systems
The University of Melbourne
Parkville, Victoria, Australia
Email: sean.maynard@unimelb.edu.au

**Atif Ahmad**

Department of Computing and Information Systems
The University of Melbourne
Parkville, Victoria, Australia
Email: atif@unimelb.edu.au

## Abstract

Research articles can support teaching by introducing the latest expert thinking on relevant topics and trends and describing practical real-world case studies to encourage discussion and analysis. However, from the point of view of the instructor, a common challenge is identifying the most suitable papers for classroom teaching amongst a very large pool of potential candidates that are not typically written for teaching purposes. Further, even in practice-oriented disciplines such as Information Security Management (ISM), high-quality journals emphasise theoretical contribution and research method rather than relevance to practice. Our review of the relevant literature did not find a comprehensive set of criteria to assist instructors in evaluating the suitability of research articles to teaching. Therefore, this research-in-progress paper presents a framework to support academics in the process of evaluating the suitability of research articles for their teaching programs.

## 1 Introduction

The selection of research articles for use in teaching by academics can be challenging. Research articles are written with a very specific scope and many articles may be required in a course, possibly even one or more articles for each topic area. This requires academics to review vast amounts of literature to identify individual research articles for their utility in teaching. This is a time consuming process, for academics that are already overloaded with administrative tasks, teaching responsibilities and research activities (Benbasat and Zmud 1999; Shkedi 1998).

The use of research articles in teaching provides a number of benefits. Research articles assist academics by showing students how to make fact-based decisions (Hemsley-Brown and Sharp 2003), can support teaching programs with "free discussion, short questioning, to improve students learning" (Abawajy 2009) and provide case studies that "promote problem solving and analysis" because "since cases are often utilized in a group setting, they provide an opportunity for students to develop teamwork, interpersonal and communications skills" (Cappel and Schwager 2002).

However, many research articles are not suitable for use in classrooms for a number of reasons. First, the academic rigour required in high quality journals impacts the usefulness of these articles in practice and teaching (Benbasat and Zmud 1999). Second, some articles are written in a manner (e.g. structure) that is simply not conducive to teaching (Lindskog et al. 1999). Third, students find some articles hard or unpleasant to read (Taylor 2007). Finally, the lifespan of research articles can be limited, especially in dynamic disciplines such as those influenced by technology (Crowley 2003).

Based on the aforementioned points discussed, this paper pursues the research question of: *"How can the suitability of research articles to information security management teaching be evaluated?"*

For the purpose of this paper we define information security management as the process of applying formal, informal and technical controls with the objective of protecting the confidentiality, integrity and availability of information in the physical and digital environment whilst maintaining strategic alignment with the organisational mission.

This paper is structured as follows. First we introduce the background to the area before describing the research methodology undertaken. We then develop a framework with categories of criteria to evaluate the suitability of research articles to a generic subject. Third, we develop a methodology using the framework so that academics can be more efficient in assessing the suitability of research articles, and subsequently show the utility of the methodology by describing a prototype application. In the discussion section we suggest how the criteria can be used in ISM to address existing deficiencies in available guidance from textbooks. Finally we conclude and offer suggestions for future work.

## 2   Background

The value of including research within teaching is widely recognised (Abawajy 2009; Benbasat and Zmud 1999; Cappel and Schwager 2002) and is often mandated by universities. Research articles often support the teaching process by promoting active learning and in-depth exploration of material (Fisher 2006; Peck 2004). However, research articles are often not written in a classroom-friendly format, or with teaching in mind. They are written for other researchers, contain technical jargon, and use complex writing styles (Lindskog et al. 1999). They may also be hard, or unpleasant to read (Taylor 2007) because they contain rigorous research approaches aimed "to establish credibility, to publish in high quality journals, to attain tenure and promotion, and to compete for research funding" (Rosemann and Vessey 2008).

Similarly, articles written in quality journals tend to lack relevance to practice. These articles focus on academic rigor over practical relevance as specified by many publication outlets (Benbasat and Zmud 1999). As a result, researchers tend to focus on explaining their rigorous research approach, making their exposure to practice-based activities infrequent and insufficient (Taylor 2007). Subsequently, research articles are less engaging for practitioners because they lack insights from real-world practice.

Furthermore, in dynamic disciplines, the lifespan of research articles decreases considerably because due to newer methods and technologies introduced continuously (Crowley 2003; Jewels et al. 2003). Therefore, there is no cumulative research tradition (Benbasat and Zmud 1999) and articles thus, become rapidly outdated. Having access to recent research is important because it is more engaging for students (Cappel and Schwager 2002) and allows them to get practice-oriented experience that will support them in the changing environment of the real world (Hsu and Blackhouse 2002).

## 3   Research Methodology

The first part of this research was to conduct a conceptual study on how to assess the suitability of research articles for teaching. We draw on guidance from Neuman (2006) on how to conduct a systematic literature review to identify articles that address this issue. Since the aim of the paper is to develop an evaluation criteria for use in Information Security Management, our approach seeks to explore literature in the related disciplines of Information Systems and Information Security literature using Google Scholar, the AIS Digital Library, and various publisher databases (e.g. Elsevier, Emerald). We searched these using a range of terms (see Figure 1).

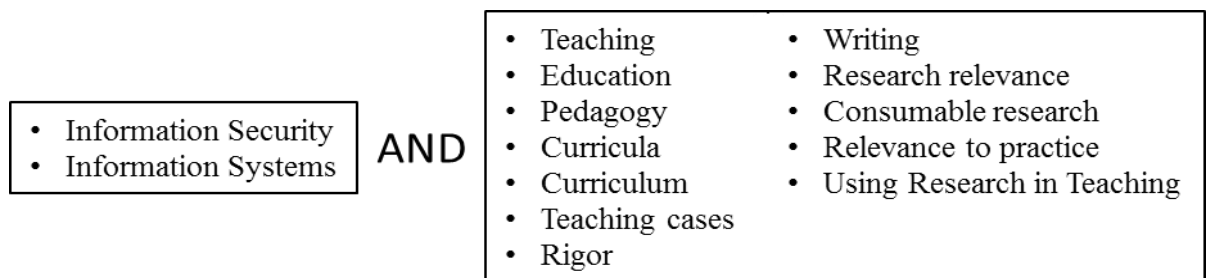

*Figure 1: Search Terms Used*

In addition to the papers identified using the search terms above, we also included those articles that were cited in these papers. Overall we identified 48 papers. Subsequently, 36 papers were discarded because they did not provide guidance or recommendations on how to make research articles and teaching cases more suitable to the student or academic community, leaving a total of twelve papers.

We analysed these papers in line with Neuman's (2006) open, axial and selective coding approach. We then developed a framework of criteria for assessing the suitability of research articles for teaching. We operationalised the framework using a multi criteria methodology as per Maynard et al. (2001). The approach determines the scores of parents (Categories) based on the scores gathered from the bottom level (Criterion). Additionally, we used the approaches in (Adelman et al. 1985) and (Goicoechea et al. 1992) to determine weightings in the two levels (Categories and Criteria) of the hierarchy.

# 4    Systematic Literature Review

Our review of the literature did not identify any articles in the IS or Information Security domains that presented a comprehensive method assessing the suitability of research articles for teaching. However, several articles presented criteria that can be used to evaluate research articles for this purpose. These criteria were aimed at the authors of research papers rather than classroom instructors. They advised on how to: (1) incorporate research into teaching; (2) increase articles' relevance to practice; and (3) write good case studies.

A number of researchers present characteristics of good research articles (Cappel and Schwager 2002; Farhoomand 2004; Kim et al. 2006). These include writing in clear and simple English (Farhoomand 2004), giving real world examples (Hackney et al. 2003), being timely (Yue 2012) and having a "hook" to motivate readership (Cappel and Schwager 2002). These publications also try to address the acknowledged problem that research articles can be unpleasant to read and lack relevance to practice.

The lack of relevance to practice in research articles makes them less accessible and less interesting to the reader. This discourages the reader from using these articles for teaching purposes (Taylor 2007). Lack of relevance to practice may be caused for a number of reasons. For instance, it is suggested that article length and complexity make it difficult for students to understand. Furthermore, because the focus of many articles is on the rigorous research approach rather than the findings (Benbasat and Zmud 1999; Rosemann and Vessey 2008) they often become hard to read as the reader gets bogged down in the detail of the rigour.

Articles used for teaching purposes need to focus more on practice-based factors around problems and topics relevant for practitioners (Rosemann and Vessey 2008). Articles can address this by providing an implementable approach to resolve practice-based problems. Furthermore, they can challenge readers' casual assumptions, paradigms or trends on practice-based areas (Benbasat and Zmud 1999). One way of achieving this is to involve practitioners during the research (Hemsley-Brown and Sharp 2003; Rosemann and Vessey 2008). Practitioners give a practice perspective to research articles by increasing the exposure researchers have to practice (Benbasat and Zmud 1999).

Teaching cases provide an alternate method of using research in teaching and are an effective tool as they allow students to develop real-life decision making, problem solving, higher-order reasoning, teamwork and communication skills (Cappel and Schwager 2002; Farhoomand 2004; Hackney et al. 2003). These skills are developed using the active learning methodology where students 'learn by doing' which is characterized as being highly motivational (Cappel and Schwager 2002).

Although teaching cases allow the development of the aforementioned skills, there is a recognised paucity of teaching cases in the Information Systems discipline (Cappel and Schwager 2002) and even more so in the information security management discipline. At the same time, the lack of cumulative research and the dynamism of this discipline cause that the few teaching cases available become rapidly outdated (Benbasat and Zmud 1999).

Literature provides guidance on how to write teaching cases for the Rotterdam School of Management (Yue 2012) and the Journal of Information Systems Education (Cappel and Schwager 2002), amongst others. Although these guides have a specific focus, they also provide general guidance for teaching case development. Kim et al. (2006) reviewed 100 teaching cases from multiple disciplines identifying strategies and core attributes of good cases. They identified five core attributes of good cases: relevant, realistic, engaging, challenging and instructional.

## 4.1    Framework to evaluate research articles for teaching purposes

From our analysis of the literature we developed a framework of criteria for the assessment of the suitability of research articles for teaching (Table 1). We analysed the literature using the three rounds of coding defined by (Neuman 2006). In the first round, open coding, we scanned the selected publications identifying recommendations on how to make research articles and teaching cases more suitable for use in classes. In the second, axial coding, we categorized the criteria according to theme. In the third round each category was divided into one or more criteria according to their specific focus.

| Category & Criteria | References |
|---|---|
| **1. Clarity**<br>1.1: How simple is the article narrative (i.e. avoiding unnecessary words, jargon, technical language, and the extended used of citations)?<br>1.2: To what extent does the article use a top down structure where the initial paragraph provides the setting and main issues of the research article? | Benbasat and Zmud (1999); Farhoomand (2004); Kavan (1998); Kim et al. (2006); Rosemann and Vessey (2008); Yue (2012) |
| **2. Succinctness**<br>2.1: To what extent does the article length match the effort required by students, as stipulated by the course, to allow them to conduct an optimal analysis of it?<br>2.2: To what extent does the article provide sufficient information to allow students to develop coherent conclusions?<br>2.3: To what extent does the article focus on the findings, rather than the inputs such as the literature review or the research methodology? | Benbasat and Zmud (1999); Cappel and Schwager (2002); Kim et al. (2006); Robey and Markus (1998); Rosemann and Vessey (2008); Senn (1998); Taylor (2007); Yue (2012) |
| **3. Objectiveness**<br>3.1: To what extent is the article written in a neutral, unbiased manner, allowing students to develop their own opinion? | Cappel and Schwager (2002); Farhoomand (2004); Robey and Markus (1998); Rosemann and Vessey (2008); Taylor (2007); Yue (2012) |
| **4. Realism**<br>4.1: To what extent does the article incorporate real world examples?<br>4.2: How authentic does the article seem given the level of evidence and facts presented?<br>4.3: To what extent does the article cite participants to increase its realism? | Benbasat and Zmud (1999); Farhoomand (2004); Hackney et al. (2003); Jewels et al. (2003); Kim et al. (2006); Rosemann and Vessey (2008); Senn (1998); Taylor (2007); Yue (2012) |
| **5. Timeliness**<br>5.1: To what extent are the research article's findings up-to-date? | Cappel and Schwager (2002); Taylor (2007); Yue (2012) |
| **6. Teaching friendliness**<br>6.1: To what extent has the article been previously assessed for use in other teaching programs? | Cappel and Schwager (2002); Kim et al. (2006); Taylor (2007) |
| **7. Depth**<br>7.1: To what extent does the article provide multiple perspectives from different stakeholders?<br>7.2: To what extent does the article provide distractors (non-pertinent features) to challenge students' analytical skills?<br>7.3: To what extent does the complexity of data (qualitative and qualitative) presented by the article help to develop students' problem solving skills?<br>7.4: To what extent does the article contain teaching aids to support student learning?<br>7.5: To what extent does the article let students make their own decisions by not providing a diagnosis of the problem?<br>7.6: To what extent does the article provide feedback on the possible actions of students?<br>7.7: To what extent does the article synthesize an existing body of research for the area of study? | Cappel and Schwager (2002); Farhoomand (2004); Kim et al. (2006); Taylor (2007); Yue (2012) |

| Category & Criteria | References |
|---|---|
| **8. Engagement**<br>8.1: To what extent does the article's storyline have a 'hook' to engage students?<br>8.2: To what extent does the article have an engaging storyline?<br>8.3: To what extent does the article include human factors such as cultural, socio-political factors, and ethical issues?<br>8.4: To what extent does the article include controversy, contrast, conflict, dilemma, or other dramatic elements?<br>8.5: To what extent does the article gradually disclose the content?<br>8.6: To what extent does the article allow students to 'learn by doing'? | Farhoomand (2004); Hackney et al. (2003); Jewels et al. (2003); Kim et al. (2006); Rosemann and Vessey (2008); Senn (1998); Taylor (2007); Yue (2012) |
| **9. Relevance to practice**<br>9.1: To what extent does the article describe current practitioner issues?<br>9.2: To what extent does the article contribute with an implementable approach to resolve a practical issue?<br>9.3: To what extent does the article stimulate a reader's casual assumptions by identifying emerging trends, structural changes or paradigms?<br>9.4: To what extent does the article reflect collaboration between researchers and practitioners? | Benbasat and Zmud (1999); Kavan (1998); Rosemann and Vessey (2008) |
| **10. Teaching objectives focus**<br>10.1: To what extent is the article applicable to the subject area?<br>10.2: To what extent does the article fit into the teaching objectives of the subject?<br>10.3: To what extent does the difficulty of the article match the ability of students in the subject? | Cappel and Schwager (2002); Jewels et al. (2003); Kim et al. (2006); Taylor (2007); Yue (2012) |
| **11. Thinking skills development**<br>11.1: To what extent does the article enable students to develop problem solving skills?<br>11.2: To what extent does the article enable students to develop critical thinking skills? | Benbasat and Zmud (1999); Farhoomand (2004); Hackney et al. (2003); Jewels et al. (2003); Kim et al. (2006) |

*Table 1: Category Framework*

## 5  Article Evaluation Methodology

The process for the evaluation of article suitability for teaching is described in this section. The process consists of two phases 'individual teaching program' and 'individual article' and within each phase are the steps required to define the weight and importance of criteria as well as to score the criteria (see Figure 2). The evaluator rates subjectively different factors on three steps (blue boxes), the rest of the steps are to make calculations to produce the final article's rating. This methodology is based on that suggested by (Adelman et al. 1985; Goicoechea et al. 1992; Maynard 1997; Maynard et al. 2001).

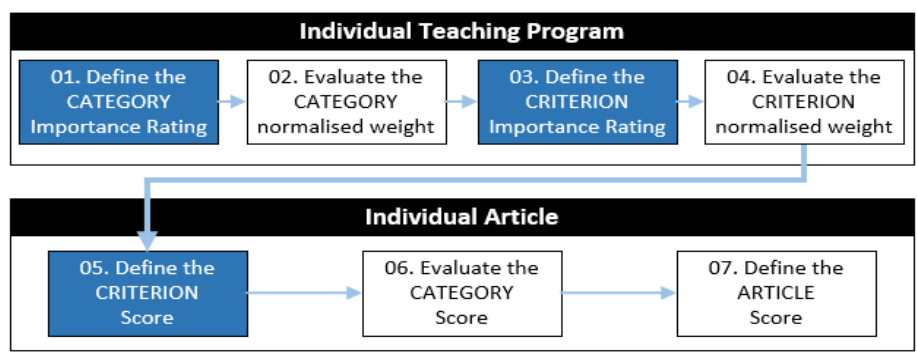

*Figure 2: Article Evaluation Methodology*

In addition, a MS Excel prototype was developed to support the process by calculating automatically the scores involved in the methodology. Thus, the evaluator is required only to introduce three values:

the category importance (*CatImp*) in Step 1, the criteria importance (*CriImp*) in Step 3, and the criteria score (*CriSco*) in Step 5 to determine article ratings.

**Step 1: Define the rating for each category considered in the evaluation process**

The evaluator rates the importance of each category (*CatImp*) to their teaching. A rating of five represents the highest importance rating; one represents the lowest importance rating; and zero represents categories that are not considered. Let us consider a scenario where an evaluator rates *Clarity* as important, whereas *Succinctness* is rated as somewhat important. For the purpose of this example the remaining criteria are rated as "not applicable" (Figure 3).

| # | Category | Category Importance | | Total |
|---|---|---|---|---|
| 01 | Clarity | 4 | Important | |
| 02 | Succinctness | 2 | Somewhat Important | |
| 03 | Objectiveness | 0 | Not Applicable | |
| 04 | Realism | 0 | Not Applicable | |
| 05 | Timeliness | 0 | Not Applicable | |
| 06 | Teaching Friendliness | 0 | Not Applicable | 6 |
| 07 | Depth | 0 | Not Applicable | |
| 08 | Engagement | 0 | Not Applicable | |
| 09 | Relevance to practice | 0 | Not Applicable | |
| 10 | Teaching objectives focus | 0 | Not Applicable | |
| 11 | Thinking Skills development | 0 | Not Applicable | |

*Figure 3: Step 1*

**Step 2: Evaluate the normalised weight of each category**

The tool determines the value of the normalised weight of each category (*CatNorWei*) by calculating the category importance (*CatImp*) divided by the sum of all the categories importance.

$$CatNorWei_i = \frac{CatImp_i}{\sum_{i=0}^{n} CatImp_i}$$

In our scenario, the tool would determine the value of the normalised weight of categories rated with non-zero values (see Figure 4).

For *Clarity*, $CatNorWei_1 = \frac{4}{6}, \ldots CatNorWei_1 = 0.67$

For *Succinctness*: $CatNorWei_2 = \frac{2}{6}, \ldots CatNorWei_2 = 0.33$

| # | Category | Category Importance | Total | Category Weight | Total |
|---|---|---|---|---|---|
| 01 | Clarity | 4 | | 66.67% | |
| 02 | Succinctness | 2 | | 33.33% | |
| 03 | Objectiveness | - | | - | |
| 04 | Realism | - | | - | |
| 05 | Timeliness | - | | - | |
| 06 | Teaching Friendliness | - | 6 | - | 100.00% |
| 07 | Depth | - | | - | |
| 08 | Engagement | - | | - | |
| 09 | Relevance to practice | - | | - | |
| 10 | Teaching objectives focus | - | | - | |
| 11 | Thinking Skills development | - | | - | |

*Figure 4: Step 2*

**Step 3: Define the rating for each criterion considered in the evaluation process**

The evaluator rates the importance of each criterion (*CriImp*) to the teaching program. A rating of five represents the highest rating; one represents the lowest rating; and zero represents criterion that are not considered. In the case of the scenario, the evaluator rates the criterion in the categories that were rated with non-zero values (Clarity and Succinctness) (Figure 5).

| # | Category | | Criterion | Criteria Importance | | Total |
|---|---|---|---|---|---|---|
| 01 | Clarity | 01.1 | How simplistic is the article narrative; avoiding unnecessary words, jargon, technical language, and the extended used of citations? | 5 | Extremely Important | 5 |
| | | 01.2 | To what extent does the article use a top down structure where the initial paragraph provides the setting and main issues of the research article? | 0 | Not Applicable | |
| 02 | Succinctness | 02.1 | To what extent does the article length match the effort required by students, as stipulated by the course, to allow them to conduct an optimal analysis of it? | 4 | Important | 9 |
| | | 02.2 | To what extent does the article provide sufficient information to allow students to develop coherent conclusions? | 5 | Extremely Important | |
| | | 02.3 | To what extent does the article focus more on the findings, rather than the inputs such as the literature review or the research methodology? | 0 | Not Applicable | |
| 03 | Objectiveness | 03.3 | To what extent is the article written in a neural, unbiased manner, allowing students to develop their own opinion? | 0 | Not Applicable | Category rated as 0 |

*Figure 5: Step 3*

**Step 4: Evaluate the normalised weight of each criterion**

The tool determines the value of the normalised weight of each criterion (*CriNorWei*) by calculating the criterion importance (*CriImp*) divided by the sum of all the criterion importance.

$$CriNorWei_{ij} = \frac{CriImp_{ij}}{\sum_{j=0}^{n} CriImp_{ij}}$$

In the scenario, the tool would determine the value of the normalised weight of both categories and criteria for those rated with non-zero values (See Figure 6).

For Clarity (1 Criterion), $\quad CriNorWei_{11} = \frac{5}{5}, \ldots CriNorWei_{11} = 1.00$ (100%)

For Succinctness (2 criteria), $\quad CriNorWei_{21} = \frac{4}{9}, \ldots CriNorWei_{21} = 0.44$ (44.44%)

$$CriNorWei_{22} = \frac{5}{9}, \ldots CriNorWei_{22} = 0.56 \text{ (55.56\%)}$$

| # | Category | | Criterion | Criterion Importance | Total | Criterion Weight |
|---|---|---|---|---|---|---|
| 01 | Clarity | 01.1 | How simplistic is the article narrative; avoiding unnecessary words, jargon, technical language, and the extended used of citations? | 5 | 5 | 100.00% |
| | | 01.2 | To what extent does the article use a top down structure where the initial paragraph provides the setting and main issues of the research article? | 0 | | 0.00% |
| 02 | Succinctness | 02.1 | To what extent does the article length match the effort required by students, as stipulated by the course, to allow them to conduct an optimal analysis of it? | 4 | 9 | 44.44% |
| | | 02.2 | To what extent does the article provide sufficient information to allow students to develop coherent conclusions? | 5 | | 55.56% |
| | | 02.3 | To what extent does the article focus more on the findings, rather than the inputs such as the literature review or the research methodology? | 0 | | 0.00% |
| 03 | Objectiveness | 03.3 | To what extent is the article written in a neural, unbiased manner, allowing students to develop their own opinion? | 0 | Category rated as 0 | - |

*Figure 6: Step 4*

**Step 5: Define to what extent the article address each criterion**

The evaluator rates the score of how well the article addresses each criterion (*CriSco*). A score of five represents the highest score; one represents the lowest score; and zero represents criterion that are not considered. In the scenario, the evaluator rates the criterion score in the categories or criteria that received non-zero importance values (Figure 7).

| | | | | | | | |
|---|---|---|---|---|---|---|---|
| 01 | Clarity | 01.1 | How simplistic is the article narrative; avoiding unnecessary words, jargon, technical language, and the extended used of citations? | 4 | To a large extent | | 80% |
| | | 01.2 | To what extent does the article use a top down structure where the initial paragraph provides the setting and main issues of the research article? | 0 | Not Applicable | | 0% |
| 02 | Succinctness | 02.1 | To what extent does the article length match the effort required by students, as stipulated by the course, to allow them to conduct an optimal analysis of it? | 5 | To a very large extent | | 100% |
| | | 02.2 | To what extent does the article provide sufficient information to allow students to develop coherent conclusions? | 2 | To a small extent | | 40% |
| | | 02.3 | To what extent does the article focus more on the findings, rather than the inputs such as the literature review or the research methodology? | 0 | Not Applicable | | 0% |
| 03 | Objectiveness | 03.3 | To what extent is the article written in a neural, unbiased manner, allowing students to develop their own opinion? | 0 | Not Applicable | | 0% Category rated as 0 |

*Figure 7: Step 5*

**Step 6: Evaluate the Articles rating in each category**

The tool determines the final score of each category (*CatSco*) by calculating the sum of each criterion score (*CriSco* from step 5) multiplied to the criterion normalized weight (*CriNorWei* step 4).

$$\textbf{\textit{CatSco}}_i = \frac{\sum_{j=0}^{n} \text{CriSco}_{ij} \times \text{CriNorWei}_{ij}}{5}$$

In the scenario, the tool would determine the value of the score of both categories that were rated in importance as different from zero (See Figure 8)

In Clarity (1 Criterion), $\textbf{\textit{CatSco}}_1 = \frac{(4 \times 1)}{5}$, … $\textbf{\textit{CatSco}}_1$ = 0.80 (80%)

In Succinctness (2 criteria), $\textbf{\textit{CatSco}}_2 = \frac{(5 \times 0.44) + (2 \times 0.56)}{5}$, … $\textbf{\textit{CatSco}}_2$ = 0.67 (66.67%)

| # | Category | | Criterion | Criterion Score | Criterion Weight | Criterion Total | Category Score |
|---|---|---|---|---|---|---|---|
| 01 | Clarity | 01.1 | How simplistic is the article narrative; avoiding unnecessary words, jargon, technical language, and the extended used of citations? | 80% | 100.00% | 80.00% | 80.00% |
|  |  | 01.2 | To what extent does the article use a top down structure where the initial paragraph provides the setting and main issues of the research article? | 0% | 0.00% | 0.00% |  |
| 02 | Succinctness | 02.1 | To what extent does the article length match the effort required by students, as stipulated by the course, to allow them to conduct an optimal analysis of it? | 100% | 44.44% | 44.44% | 66.67% |
|  |  | 02.2 | To what extent does the article provide sufficient information to allow students to develop coherent conclusions? | 40% | 55.56% | 22.22% |  |
|  |  | 02.3 | To what extent does the article focus more on the findings, rather than the inputs such as the literature review or the research methodology? | 0% | 0.00% | 0.00% |  |
| 03 | Objectiveness | 03.3 | To what extent is the article written in a neural, unbiased manner, allowing students to develop their own opinion? | 0% | - | - | 0.00% |

*Figure 8: Step 6*

**Step 7: Evaluate the Article's Rating**

The tool determines the final rating of the article (*ArtRat*) by calculating the sum of each category rating (*CatSco* from Step 6) multiplied to the category normalized weight (*CatNorWei* Step 2).

$$\textbf{\textit{ArtRat}} = \sum_{i=0}^{n} CatSco_i \times CatNorWei_i$$

From the Step 6 Example: the tool determines the final rating of the article by calculating the sum of each category rating multiplied by their weight (Figure 9). Considering only non-zero rated categories (2 Categories), **ArtRat** = (0.80 × 0.67) + (0.67 × 0.33), … **ArtRat** = 0.7556, … ArtRat = 75.56 %

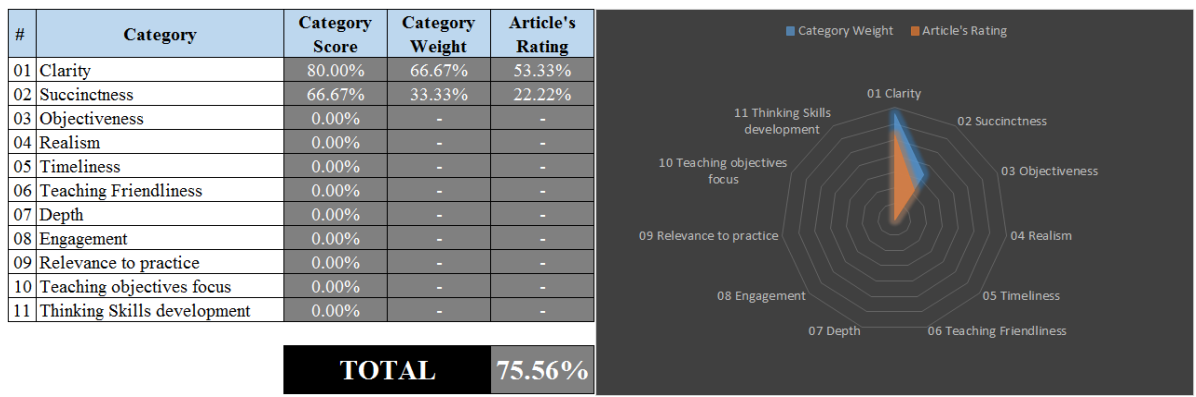

*Figure 9: Step 7*

This example shows the complete process of the two mentioned parts of the methodology. First, rating the importance of the categories and criteria for the evaluator's teaching program. Second, rating one article based on the first part. In a real case example, the evaluator would rate more articles to compare their ratings. To get the total score of another article, the evaluator needs to repeat the methodology from Step 5.

# 6   Discussion: The Utility of the 'Suitability of Research Paper Criteria' to Information Security Management

In this section we suggest how the suitability criteria can be used to address the gaps in ISM guidance from available textbooks. In order to do this we focus on the criteria related to suitability to teaching topics (categories 9, 10 and 11), rather than the criteria aimed at evaluating the suitability to classroom teaching.

## 6.1 Relevance to Practice across Industries and Contexts (Category 9)

ISM instructors may be interested in exposing students to ISM practices in a range of contexts. For example, relating to industry sectors ISM plays a key role in: (1) Critical Infrastructure Protection (CIP) from National Security threats (see Theoharidou et al. (2007) for an argument for CIP to be included in curricula and Beraud and Ahmad (2011)) for a discussion of why risk methods should consider CIP); (2) protecting private enterprise from local and international competitors (see a range of ISM controls to protect competitive advantage in Ahmad et al. (2014a); and (3) privacy of personnel information in public organisations (Bélanger and Crossler 2011). Further, security issues in Small to Medium-sized Enterprise (SMEs) are frequently different to that of larger organisations (Barlette and Fomin 2008; Ng et al. 2013). ISM is strongly influenced by differentiators among employees such as national culture (Ifinedo 2009) and behavioural archetypes (Crossler et al. 2013). ISM is also influenced by organisational differentiators such as organisational culture (Lim et al. 2010; Lim et al. 2009) and governance (Koh et al. 2005). All of these issues are critical for ISM practitioners to consider when developing effective security strategy in organisations.

ISM instructors may also look at deficiencies in the way organisations implement security guidance e.g. from industry standards. For example, studies have pointed out deficiencies in the implementation of information security risk assessment (ISRA) methods (e.g. see Shedden et al. (2010a). Similarly, there have been a number of studies looking at deficiencies in the incident response process (Ahmad et al. 2012; Ahmad et al. 2015; Tøndel et al. 2014). Although these case studies are hard to find because organisations rarely give access to their sensitive information and functions (see Kotulic and Clark (2004) and Tøndel et al. (2014)), however they provide valuable insights for students that relate to real-world ISM challenges.

Instructors will find that the discussion of ISM in most textbooks tends to take a narrow view of the range of formal and informal controls that fall under ISM (see Dhillon (2006) for a discussion on the distinctions between formal, informal and technical controls and Whitman and Mattord (2014) as an example of a management-oriented textbook that covers a range of security controls).

First, the emphasis continues to remain on traditional controls such as Policy, Risk and SETA whilst neglecting other critical areas such as intra-organisational liaison (communication, collaboration and coordination) between ISM and other parts of the organisation (Alshaikh et al. 2014), as well as the core security strategy process (see Baskerville and Dhillon (2008)). It is also unclear whether ISM should include Digital Forensic Readiness (see Elyas et al. (2015) for a management perspective on Digital Forensics and commentary on security contributions). Further, the discussion of managerial activities remains at a high-level, which does not provide enough detailed guidance for organisations seeking to implement the functions internally (Alshaikh et al. 2014).

## 6.2 Teaching Objectives: Imparting the Management Perspective of Information Security (Category 10)

The primary teaching objectives for instructors in Information Security Management is how to prepare students for a career in the discipline by providing: (1) an understanding of the management perspective of Information Security; and (2) access to knowledge that is relevant to real-world practice across a range of industries and contexts where ISM may be applied (see 6.2) (Ahmad and Maynard 2014; Martini and Choo 2014).

Regarding the first objective, ISM instructors will struggle to impart an authentic management perspective of Information Security without falling into the conventional 'IT Security' discourse typical in ISM textbooks. The ISM instructor can find a number of recent papers that depart from the traditional view of information assets as being discrete, enumerated and situated in the formal business process (e.g. see Shedden et al. (2009); Shedden et al. (2011); Shedden et al. (2010b) for a distributed cognitive view of information within informal business practice). Further, a number of recent studies focus on the security of 'tacit knowledge' in 'human containers' (see Ahmad et al. (2014a); Manhart and Thalmann (2015). These papers espouse the idea that enterprise security must adopt an 'information-centric' rather than 'IT-centric' view (see Ahmad and Ruighaver (2005); Winkler (1996)).

### 6.3   Thinking skills to support ISM practice (Category 11)

A key topic that has been largely neglected is how ISM practitioners should 'strategize' by leveraging their resources to best advantage to address security risk. Security strategy literature has pointed out that ISM managers facing unpredictable and transient threats in the shape of intelligent adversaries (e.g. in cases of industrial espionage and cyber terrorism) must adopt a 'warfare' mindset (this view was first presented in Baskerville (2005) and then tested in Baskerville et al. (2014) which implies security situation awareness (see Webb et al. (2014) for a model of situation awareness) must be developed in combination with tactical speed and agility.

Some research has looked at the range of strategic and tactical paradigms that can be implemented in organisations (e.g. see Tirenin and Faatz (1999) and Ahmad et al. (2014b)) however there has not been much discussion on the particular thinking skills such as game-theoretic approaches needed to employ a combination of strategies effectively especially in asymmetric situations (e.g. see game-theoretic approaches in Cavusoglu et al. (2008), and discussion of asymmetry in cyber-physical situations in Ahmad (2010)).

## 7   Conclusion

This paper has identified criteria for the evaluation of the applicability of research articles for use in teaching and has operationalised these criteria into a methodology for article assessment. This methodology enables academics evaluating, in seven steps, one or more articles to determine their suitability for teaching use. It allows comparisons of papers to occur to enable the choice of the most suitable papers for teaching programs.

We consider that the framework and methodology provided answers the proposed research question: ***"How can research articles applicability be evaluated for use in teaching?"*** The framework and methodology can be used to evaluate research articles suitability for teaching. The framework provides a set of comprehensive categories to be considered in the assessment of research articles applicability to teaching. The criteria, formatted as questions, provided evaluators with a set of requirement to meet in the research articles. Finally, the methodology allows evaluators to follow a descriptive process to complete this task.

The importance of the criteria is linked to the discipline of the teaching program. For example, we consider that in practice-base disciplines the category "relevance to practice" is going to be a key category. The reason is that practice-oriented issues are important for this type of discipline and they are not sufficiently discussed in literature.

This paper contributes to theory as it has analysed the literature to identify 11 areas and 33 criteria to analyse papers with the purpose of selecting papers to use in teaching. This is the first study, that the authors are aware of that has analysed and synthesised this literature to from an analysis framework which is then used by an evaluation methodology.

The paper also has contributed to practice, especially in the area of paper assessment by academics for teaching purposes. Using the criteria and methodology proposed will enable academics to make quick decisions on the suitability of teaching cases and research papers to be used in teaching. This will reduce the time required for the selection of articles.

This paper has focused mainly on the information systems domain in its search for literature. There may be research conducted in other domains about incorporation of research papers into teaching. Future work will identify these and incorporate them into our framework and methodology. The methodology has limitations as it relies on unbiased assessment of articles and required self-assessment of the criteria by academics. This requires that academics read the articles very carefully and that they are very well-versed in the articles reviewed.

There are a number of future projects that build on this work. First, the applicability of text books to the classroom environment may be able to be assessed. Second, a research project is underway to assess the literature in the Information Security Domain using this tool, with the aim of selecting relevant articles for a Masters level Information Security Management course that the authors teach.